\documentclass[epj]{svjour}
\usepackage{amsmath}
\usepackage{mathtools}
\usepackage{algorithm}
\usepackage{algorithmic}
\usepackage{graphics}
\usepackage{amssymb}
\usepackage{mathtools,cuted}
\usepackage{color}
\usepackage{lipsum}
\usepackage[square, comma, sort&compress, numbers]{natbib}
\usepackage{dcolumn}
\usepackage[figuresright]{rotating}
\usepackage[colorlinks=true,linkcolor=blue, citecolor=blue]{hyperref}

\begin{document}
\title{Community detection by label propagation with compression of flow}

\author{Jihui Han\thanks{\email{jh@mails.ccnu.edu.cn}} \and Wei Li\thanks{\email{liw@mail.ccnu.edu.cn}} \and Zhu Su \and Longfeng Zhao \and Weibing Deng
}

\institute{Complexity Science Center, Institute of Particle Physics, Central China Normal University, Wuhan 430079, China}
%
%
\abstract{
 The label propagation algorithm (LPA) has been proved to be a fast and effective method for detecting communities in large complex networks. However, its performance is subject to the non-stable and trivial solutions of the problem. In this paper, we propose a modified label propagation algorithm LPAf to efficiently detect community structures in networks. Instead of the majority voting rule of the basic LPA, LPAf updates the label of a node by considering the compression of a description of random walks on a network. A multi-step greedy agglomerative strategy is employed to enable LPAf to escape the local optimum. Furthermore, an incomplete update condition is also adopted to speed up the convergence. Experimental results on both synthetic and real-world networks confirm the effectiveness of our algorithm.
} 
\maketitle
\section{Introduction}
\label{intro}
Real-life complex systems in many research fields such as biology, sociology, economy and computer science, can be studied as networks with nodes representing for individuals and links for interactions or relations between individuals. Many networks exhibit the so-called community structure: nodes tend to organize themselves in groups such that connections are denser within groups while sparser between groups. Community structure is a prominent feature of complex networks, as it often represents functional modules with nodes of common properties and accounts for the functionality of the system. Community detection enables us to probe the organization and functional behavior of real-world systems, therefore has been paid much attention and applied to many kinds of networks, including the collaboration networks \cite{Newman2001}, social networks \cite{Scott2000}, and biological networks \cite{Fell2000}, etc.

Community detection has been studied as the graph partitioning in computer science for decades and remains quite challenging. Algorithms to detect reasonably good quality communities have been proposed and improved extensively \cite{Fortunato201075}, especially in recent years, such as Girvan-Newman algorithm \cite{PhysRevE.69.026113}, spectral clustering \cite{PhysRevE.74.036104,White05}, multi-state spin model \cite{PhysRevLett.93.218701,Son2006,Kumpula2007} (e.g., q-state Potts model), random walk \cite{Bubak2004,Pons2005,Ochab2013}, modularity optimization \cite{1742-5468-2008-10-P10008,Barber2013,Waltman2013,Xiang2012} and statistical inference \cite{Newman05062007,1742-5468-2010-04-P04028,PhysRevE.79.036111,PhysRevLett.100.258701}.

As one of the fastest algorithms for community detection, the label propagation algorithm (LPA) \cite{PhysRevE.76.036106} uses the network structure alone to guide its process and requires neither parameters nor optimization of any object function. It starts by assigning each node a unique label, indicating the community it belongs to. At every label propagation step, each node sequentially updates its label to a new one that most of its neighbors own. If more than one label is the most frequent, the new label is chosen randomly among them. The label propagation step is performed iteratively until each node has a label that is the most frequent among its neighbors'. Through this iterative process, the densely connected groups of nodes form consensus on one label to form communities. Finally, LPA converges when no node changes its label anymore. Therefore, nodes with the same label are classified into the same community. In addition to its nearly linear time complexity, LPA introduces no parameter and requires no priori information of communities, and thus is suitable to process large-scale networks with millions of nodes and edges.

Due to the frequent tie-breaks and the random order update strategy, LPA usually delivers multiple partitions starting from the same initial condition, with different random seeds. Raghavan et al. \cite{PhysRevE.76.036106} proposed to label each node with the set of all labels it has in different partitions to detect possible overlapping communities. However, in a recent paper, Tibely and Kertesz \cite{Tibély20084982} showed that this method was equivalent to finding the local minima in a simple zero-temperature kinetic Potts model. The number of such local minima was found to be much larger than the number of nodes in the underlying network. Aggregating partitions suggested by Raghavan et al. \cite{PhysRevE.76.036106} leads to a fragmentation of the resulting partition in small clusters when the number of aggregated partitions is large.

In order to eliminate undesired solutions, Barber and Clark \cite{PhysRevE.80.026129} proposed a modularity-specialized LPA (LPAm) to constrain the label propagation process, which is inclined to get stuck in poor local maximum of modularity. To solve this problem, Liu et al. \cite{Liu20101493} introduced an advanced modularity-specialized LPA (LPAm+), which is more stable than LPAm. Due to the usage of modularity, the capability of both algorithms will be affected by the resolution limit \cite{Fortunato02012007}.

Leung et al. \cite{PhysRevE.79.066107} have found that LPA often yields partitions with one giant community together with much smaller ones when applied to online social networks. In order to avoid such a disturbing feature, they proposed a modified method by adding a decreasing score assignment for each label in label propagation process (LPA-$\delta$), which encourages the formation of a stronger local community and deters the occurrence of trivial solutions. Tests of LPA-$\delta$ on the LFR benchmark produced good results \cite{PhysRevE.78.046110}. To save the running time of LPA-$\delta$, Leung et al. proposed to avoid label update of those nodes with high neighbor purity \cite{PhysRevE.79.066107}. Since the neighbor purity ignores contribution of the small degree nodes to the community detection, the detection precision is not high enough.

In this paper, we propose the LPAf which introduces a new update rule to update the label of a node by taking into account the compression of flow (random walks on a network), and uses an incomplete update condition in label propagation process to speed up the convergence. Like LPAm+, LPAf employs a multi-step greedy agglomerative algorithm (MSG) \cite{Oades2008} to simultaneously merge multiple pairs of communities. Although LPAf is also applicable to weighted and directed networks, we currently focus on unweighted and undirected networks. The paper is organized as follows. In Sec.~\ref{lpaf}, we present our new method in detail. Experimental results on synthetic and real-world networks are shown in Sec.~\ref{results}. Finally, the main findings are summarized in Sec.~\ref{conclu}.

\section{Algorithm}
\label{lpaf}
To reveal community structures in networks, Rosvall and Bergstrom \cite{Rosvall2009} introduced an information theoretic approach (known as Infomap algorithm). They use the probability flow of random walks on a network as a proxy for information flows in real systems and decompose the network into communities by compressing a description of the probability flow.

For a network partition $C$ of $n$ nodes containing $c$ communities, the average description length of random walks is defined as \cite{Rosvall2009},
\begin{equation}
\label{mapequation}
L(C) = q_\curvearrowright H(\Omega) + \sum_{i=1}^c p_{i\circlearrowright} H(P^i) ,
\end{equation}
where
\begin{equation}
\label{submapequation1}
H(\Omega) = -\sum_{i=1}^c \frac{q_{i\curvearrowright}}{q_\curvearrowright}\log\left(\frac{q_{i\curvearrowright}}{q_\curvearrowright}\right),
\end{equation}
and
\begin{equation}
\label{submapequation2}
H(P^i) = -\frac{q_{i\curvearrowright}}{p_{i\circlearrowright}} \log \left(\frac{q_{i\curvearrowright}}{p_{i\circlearrowright}}\right) -\sum_{\alpha\in i}\frac{p_\alpha}{p_{i\circlearrowright}}\log\left(\frac{p_\alpha}{p_{i\circlearrowright}}\right),
\end{equation}
in which $\alpha=1,2,\dots,n$ and $i=1,2,\dots,c$.

Here $q_{i\curvearrowright}$ is the probability of exiting community $i$, $q_\curvearrowright = \sum_{i=1}^c q_{i\curvearrowright}$ is the probability that the random walker switches to a different community at any given time step, $p_\alpha$ is the probability of visiting node $\alpha$ and $p_{i\circlearrowright} = \sum_{\alpha\in i}p_\alpha +q_{i\curvearrowright}$ is the fraction of time the random walker spends in community $i$ plus the probability of exiting that community.

By combining Eqs.~(\ref{mapequation}), (\ref{submapequation1}) and (\ref{submapequation2}), the expanded form of map equation can be written as,
\begin{equation}
\begin{split}
L(C)=&q_\curvearrowright \log \left(q_\curvearrowright\right) - 2\sum_{i=1}^{c}q_{i\curvearrowright} \log\left(q_{i\curvearrowright}\right) \\ &-\sum_{\alpha=1}^{n}p_\alpha\log\left(p_\alpha\right) + \sum_{i=1}^{c}p_{i\circlearrowright} \log \left(p_{i\circlearrowright}\right).
\label{expandedmapequation}
\end{split}
\end{equation}
Note that the term $\sum_1^n p_\alpha\log p_\alpha$ is independent of partitioning. Consequently, when we update the label of node $\alpha$ from $i$ to $j$, it is sufficient to only keep track of changes of $q_{i\curvearrowright}$ and $q_{j\curvearrowright}$. They can be easily derived for any update event, and updating them is fast and straightforward (see Appendix A for details).

\begin{figure}
	\resizebox{0.48\textwidth}{!}{%
		\includegraphics{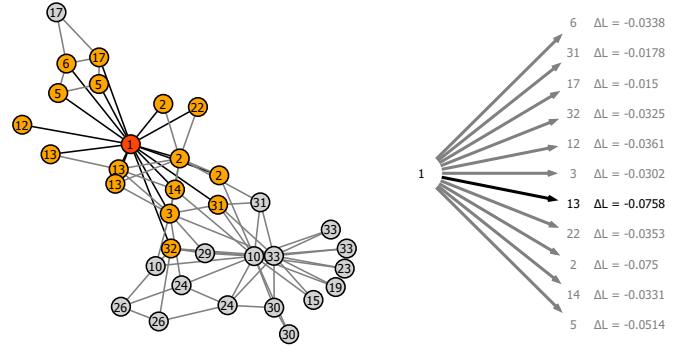}
	}
	\caption{(Color online) Snapshot of a label propagation step. Labels represent communities that nodes belong to. The node to be updated and its neighbors are orange and light orange respectively. Changes in average description length are shown on the right panel. The minimum of $\Delta L$ is highlighted by dark gray. Thus, according to our new update rule, the node ought to change its label from 1 to 13 in this case.}
	\label{karate1}       
\end{figure}

We extend the LPA by modifying the label update rule so that the average description length $L(C)$ can be minimized. When update the label for $\alpha$, we pick the one with the smallest $\Delta L$ (as illustrated on \emph{karate} network in Fig.~\ref{karate1}). Hence, our new update rule can be expressed as,	
\begin{equation}
\label{updaterule}
l_\alpha^{new} = \arg\min_{l\in N_l(\alpha)}\Delta L\left(\alpha, l_\alpha, l\right),
\end{equation}

\noindent where $l_\alpha$ is the current label for node $\alpha$, $l_\alpha^{new}$ is the new label for node $\alpha$, $N_l(\alpha)$ includes the labels of the neighboring nodes of $\alpha$, $\Delta L(\alpha, i, j)$ is the change of $L$ when update the label of node $\alpha$ from community $i$ to $j$ (see Appendix A for details), and $\arg\min_l$ returns the label $l$ that minimizes $\Delta L$. If more than one label shares the same minimum of $\Delta L$, the new label is chosen randomly among them. The label propagation step is performed iteratively until $L$ no longer decreases.

In our tests, this update rule helps form local subgroups. However, it alone does not provide satisfying performance in dealing with large-scale networks, as it usually gets stuck in poor local minima in $L$ space.

\begin{figure}
	\resizebox{0.48\textwidth}{!}{%
		\includegraphics{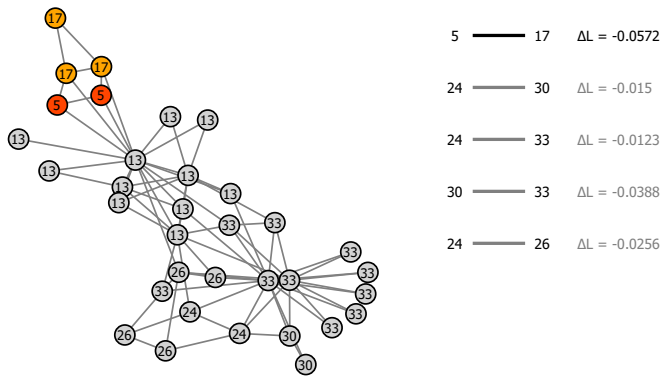}
	}
	\caption{(Color online) Snapshot of a merging event. Labels represent communities that nodes belong to. Changes in average description length for merging pairs of communities are shown on the right panel. In this case, communities 5 and 17 should be merged firstly as the merging of them leads to the smallest $\Delta L$, followed by 30 and 33, 24 and 26. Community pairs 24 and 30, 24 and 33 are excluded because community 24 has already been merged before.}
	\label{karate37}       
\end{figure}

In order to escape the local minimum, we adopted a greedy rule for merging communities that minimizes $L$, i.e., when the LPA with our new update rule gets stuck in a local minimum (no decrease in $L$ can be achieved via further label propagation), we calculate the changes of $L$ for merging pairs of communities, and merge those pairs that decrease $L$ the most. In actual operation, we employ the MSG technique to simultaneously merge multiple pairs of communities (as illustrated in Fig.~\ref{karate37}). After merging communities, we escape the local minimum. Then we should perform another round of label propagation using the new update rule. This is analogous to downhill into another local minimum. However, it is not guaranteed that the new local minimum reached is good enough. Hence the above process should be repeated indefinitely until $L$ no longer decreases.

To avoid unnecessary updates in each iteration of LPAf, the incomplete update condition proposed in Ref.~\cite{6004645} was adopted. Consequently, we only update the labels of the active nodes which would change their labels if they attempt to update. A list containing all currently active nodes is maintained to allow the algorithm to finish execution when the list is empty (i.e., we only track the nodes that potentially change their labels). The pseudo-code of our algorithm is presented in Algorithm~\ref{code:LPAf}.

\begin{algorithm}[h]
	\caption{LPAf}
	\begin{algorithmic}[1]
		\STATE{each node is assigned a unique label}
		\STATE{perform label propagation using our new update rule}
		\WHILE {$\exists$ community pairs with $\Delta L<0$}
		\STATE merge those community pairs using the MSG;
		\STATE{perform label propagation using our new update rule};
		\ENDWHILE
	\end{algorithmic}
	\label{code:LPAf}
\end{algorithm}

\section{Results}
\label{results}
Many metrics have been proposed to quantify the quality of a network partition. When the ground truth is unknown, a common measure for the significance of the identified community structure is \emph{modularity} $Q$ \cite{PhysRevE.69.026113}, which is defined as,
\begin{equation}
Q = \frac{1}{2m}\sum_{u,v=1}^n\left(A_{uv}-P_{uv}\right)\delta\left(l_u, l_v\right),
\label{modularity}
\end{equation}
where $m$ is the total number of edges in the network. $A_{uv}=1$ if nodes $u$ and $v$ are connected and 0 otherwise, $P_{uv}=k_uk_v/2m$ is the probability in the null model that an edge exists between nodes $u$ and $v$, and $\delta\left(i,j\right)$ is the Kronecker function: two vertices $u$ and $v$ provide a non-zero contribution to the value of $Q$ if and only if they belong to the same community. The concept of \emph{modularity} is based on the idea that a random graph is not expected to exhibit the community structure.

For a more sufficient assessment of the significance of detected communities, we also adopt the \emph{modularity density} $Q_{ds}$ \cite{modularitydensity} and the \emph{conductance} $\Phi$ \cite{bollobas1998modern} metrics.

Given an undirected network, the modularity density is defined as
\begin{equation}
\begin{aligned}
Q_{ds} = \sum_{c_i\in C}\left[\frac{|E_{c_i}^{in}|}{m}d_{c_i}-\left(\frac{2|E_{c_i}^{in}|+|E_{c_i}^{out|}}{2m}\right)^2 \right.\\
\left. -\sum_{c_j\in C,c_j\ne c_i}\frac{|E_{c_i,c_j}|}{2m}d_{c_i,c_j}\right],
\end{aligned}
\label{modularity_density}
\end{equation}
where $C$ is the set of all the communities, $c_i$ is any given community in $C$, $d_{c_i}=\frac{2|E_{c_i}^{in}|}{|c_i|\left(|c_i|-1\right)}$ is the internal density of community $c_i$, $d_{c_i,c_j}=\frac{|E_{c_i,c_j}|}{|c_i||c_j|}$ is the pair-wise density between communities $c_i$ and $c_j$, $|E_{c_i}^{in}|$ is the number of edges between nodes within community $c_i$, $|E_{c_i}^{out}|$ is the number of edges from the nodes in community $c_i$ to the nodes of other communities, and $|E_{c_i,c_j}|$ is the number of edges between communities $c_i$ and $c_j$. Compared to modularity, the \emph{modularity density} is an improved measurement for assessing the quality of communities, since it does not suffer from the well-known resolution limit of modularity.

For a community $c_i$, the conductance is defined as
\begin{equation}
\Phi(c_i) = \frac{\sum_{u\in c_i,v\not\in c_i}A_{uv}}{\sum_{u\in c_i}k_u},
\label{conductance}
\end{equation}
where $k_u$ is the degree of node $u$. Informally, \emph{conductance} is the fraction of total edge volume that points outside the community $c_i$. Lower values of \emph{conductance} imply that the communities have more internal connections than external ones, and thus represent more significant communities. Due to the fact that conductance cannot be easily extended to an entire community structure of a network, results are commonly assessed at different scales separately in the form of \emph{network community profile (NCP)} \cite{leskovec2009community} plots.

For networks with known community structures, two metrics from the field of information theory \cite{MacKay03a} are adopted to compare identified communities with the true ones. The first one, \emph{normalized mutual information (NMI)} \cite{1742-5468-2005-09-P09008}, estimates the amount of information correctly extracted by the detection algorithms and has become a de facto standard to quantify the quality of a detected partition with respect to the ground truth. It is defined as,
\begin{equation}
NMI(X,Y) = \frac{2I\left(X,Y\right)}{H\left(X\right)+H\left(Y\right)},
\label{NMI}
\end{equation}
where $X$ and $Y$ denote two partitions of the network, $I\left(X,Y\right)=H\left(X\right)-H\left(X|Y\right)$, $H\left(X\right)$ is the Shannon entropy of $X$ and $H\left(X|Y\right)$ is the conditional entropy of $X$ given $Y$. \emph{NMI} equals 1 if the detected partition is identical to the real one, whereas it has an expected value of 0 if the detected partition is totally independent of the real one.

The second metric is the \emph{variation of information(VOI)} \cite{MEILA2007873}, which has several desirable properties with respect to \emph{NMI}. Specifically, it can be regarded as a kind of distance in the space of partitions. \emph{VOI} of $X$ and $Y$ is defined as
\begin{equation}
VOI(X,Y)=H(X|Y)+H(Y|X).
\label{VOI}
\end{equation}
Thus, lower values represent higher similarities between partitions. The value of \emph{VOI} ranges from 0 to $\log N$, where $N$ is the network size. Therefore, we divide the obtained values by $\log N$ for meaningful comparisons.

We have tested our algorithm on both synthetic and real-world networks. For comparisons, five algorithms, the original LPA \cite{PhysRevE.76.036106}, the neighbor strength driven LPA (nsdLPA) \cite{6004645}, the Louvain method \cite{1742-5468-2008-10-P10008}, the Infomap algorithm \cite{Rosvall2009}, and the fine-tuned modularity density algorithm (FineTune) \cite{FineTune}, are included in the experiments as references. The nsdLPA enhances the basic LPA by taking into account the positive neighborhood strength, and is generally efficient in practice \cite{6004645}. The Louvain method is a greedy optimization algorithm that attempts to optimize the modularity of a partition, which usually produces high modularity values and is by far one of the most widely used method for detecting communities in large networks \cite{1742-5468-2008-10-P10008}. The Infomap algorithm decomposes a network into communities by compressing a description of information flow on the network as mentioned above \cite{Rosvall2009}. The FineTune algorithm iteratively attempts to improve the modularity density measurement by splitting and merging the given network community structure \cite{FineTune}.

\subsection{Tests on synthetic networks}
\label{synthetic_results}
We first tested our method on the well-known GN benchmark \cite{Girvan11062002}, and compared the results to the counterparts of other methods. The GN benchmark network consists of 128 nodes, each with expected degree 16, which are divided into four groups with 32 nodes each. The mixing parameter $\mu$ measures the ratio of the external degree of a node with respect to its community to the total degree of the node.

\begin{figure}
	\centering
	\resizebox{0.48\textwidth}{!}{%
		\includegraphics{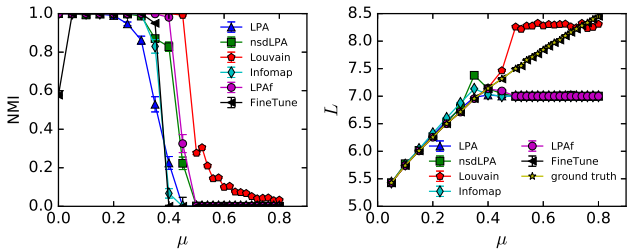}
	}
	\caption{(Color online) Average NMI and $L$ for different algorithms as a function of $\mu$ on the GN benchmark networks. Each data point is computed over 100 different network realizations.}
	\label{GNbench}
\end{figure}

The results of different methods on the GN benchmark networks are shown in Fig.~\ref{GNbench}. As can be seen, Louvain method performs fairly well on the GN benchmark network. This indicates that the community size of the GN benchmark network is not below the resolution limit, and the optimization of modularity indeed reveals the true partitions. LPAf performs next to Louvain method, and significantly better than the rest four methods. All the methods except Louvian and FineTune arrive at the same stable value of $L$ at high $\mu$, which corresponds to the trivial partition. LPAf cannot detect the real communities in this range by minimizing $L$, because the trivial partition has a lower $L$ than the real partition.

We also adopted the LFR benchmark \cite{PhysRevE.78.046110}, which is a special case of the planted $l-$partition model \cite{RSA:RSA1001}. LFR networks are similar to real-world networks, since all of them are characterized by heterogeneous distributions of node degrees and community sizes. In our experiments, the parameters are fixed as follows: node degrees and community sizes are governed by the power law, with exponents being -2 and -1 respectively; the maximum degree is 50; the ranges of community sizes are [10,50] and [20,100] for smaller and bigger communities respectively; the network size $N$ is either 1000 or 5000. The significance of community structure is controlled by a mixing parameter $\mu\in [0,1]$ where smaller values correspond to more obvious community structure. $\mu$ is the expected fraction of links of a node connecting to other communities.

\begin{figure}
	\centering
	\resizebox{0.48\textwidth}{!}{%
		\includegraphics{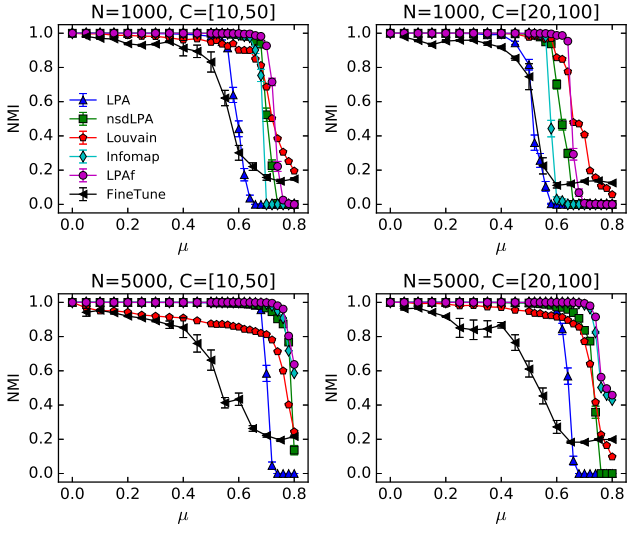}
	}
	\caption{(Color online) Average NMI for different algorithms as a function of $\mu$ on LFR networks. The number of vertices is either 1000 (small scale) or 5000 (large scale), and the ranges of community sizes are [10, 50] (smaller community, left panel) and [20,100] (larger community, right panel). Each data point is averaged over 100 different network realizations.}
	\label{nmi}
\end{figure}

Results are assessed in terms of average NMI, shown in Fig.~\ref{nmi}, which shows that, the LPAf outperforms other methods consistently for a wide range of $\mu$. In contrast to the GN benchmark, Louvain method fails to detect the real communities even when $\mu$ is small for larger networks with smaller communities. This is due to the well-known resolution limit of modularity, i.e., there exists a size cutoff below which modularity cannot identify communities \cite{Fortunato02012007}. In order to optimize modularity, Louvain method tends to merge natural communities into much larger ones, which leads to rather poor performance. FineTune does not have remarkable performance either, as it also starts to fail for low values of $\mu$. The nsdLPA performs better than LPA due to the consideration of the positive neighbor strength. Infomap performs comparably with LPAf in larger networks but is outperformed by LPAf when networks are small. Moreover, LPAf is more stable than LPA, because of the lower standard deviation of its NMI scores. The results thus confirm that LPAf performs better than or at least as well as the rest five methods in all the LFR networks.

\begin{figure}
	\centering
	\resizebox{0.48\textwidth}{!}{%
		\includegraphics{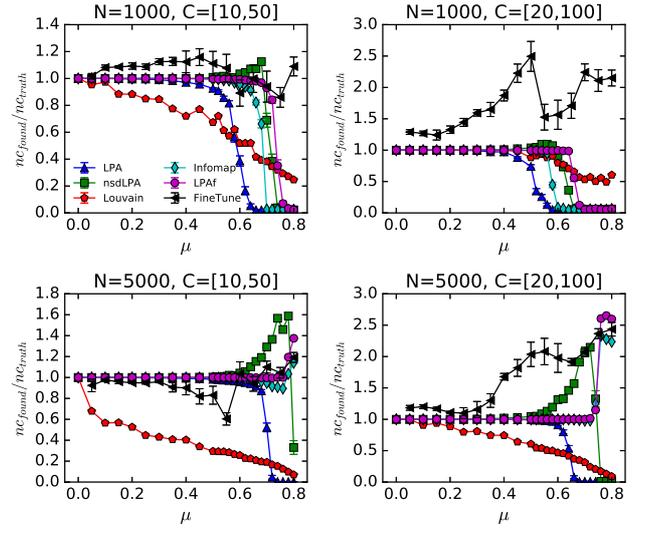}
	}
	\caption{(Color online) Average ratio of the number of detected communities to the number of actual communities for different algorithms with varying $\mu$ on LFR networks. Each data point is averaged over 100 different network realizations.}
	\label{nc}
\end{figure}

To further address the validity of LPAf, we also computed the average ratio of the number of detected communities to the number of actual ones and showed them in Fig.~\ref{nc}. As can be seen, the number of communities detected by the LPAf is very close to the actual one up to a high $\mu$ in all cases. The number of communities detected by nsdLPA is larger than the actual one at high values of $\mu$, which implies that nsdLPA tends to form local subgroups and favors smaller communities due to the consideration of neighborhood strength. Louvain method tends to find less communities than planted ones due to the resolution limit of modularity, whereas FineTune normally detects more communities than actual ones in most cases. This indicates that FineTune resolves, to a certain degree, the resolution problem of Louvain method. In most cases, Infomap tends to find sightly less communities than actual ones.

\begin{figure}
	\centering
	\resizebox{0.48\textwidth}{!}{%
		\includegraphics{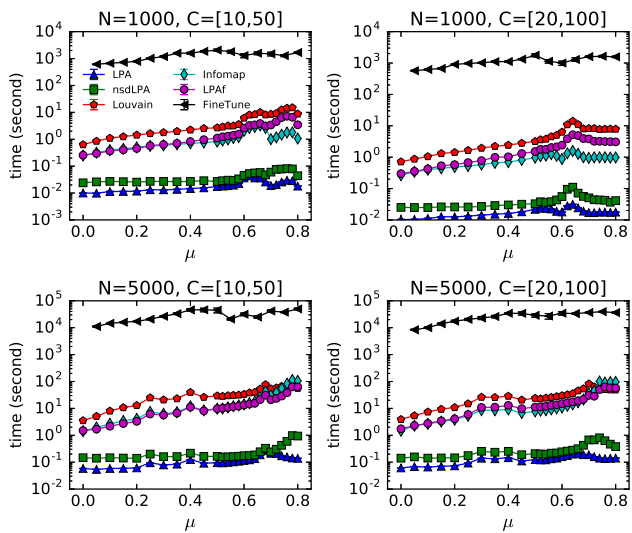}
	}
	\caption{(Color online) (Color online) Average elapsed time as a function of $\mu$ of different algorithms on the LFR networks. Each data point is an average over 100 different network realizations.}
	\label{time}
\end{figure}

To compare the computational loads of different methods, we plot the average elapsed times in Fig.~\ref{time}. Generally, the running times of all methods increase when $\mu$ gets larger. This is due to that when $\mu$ is small, the communities are well separated and all the methods can easily detect them in a short period of time. When $\mu$ increases to a specific value where the community structure still persists but is much more difficult to be revealed, the convergence speed slows down and thus results in peaks of the curves. When $\mu$ continues increasing, most of the methods cannot detect non-trivial communities and converge sightly faster than at the transition stage. Specifically, LPA and nsdLPA are faster than the rest four algorithms. LPAf, Louvain and Infomap exhibit similar time consuming patterns.

\begin{figure}
	\centering
	\resizebox{0.48\textwidth}{!}{%
		\includegraphics{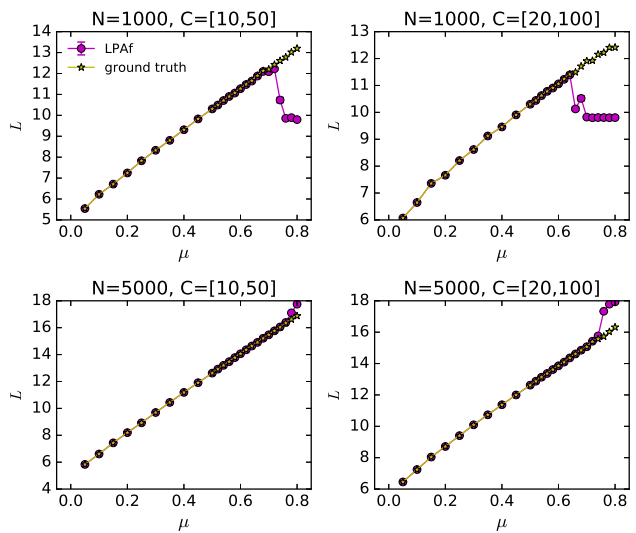}
	}
	\caption{(Color online) The comparison of $L$-values between detected and real partitions.}
	\label{L_LPAf_GT}
\end{figure}

To test how well LPAf performs in finding the local minimum in $L$ space, we computed the values of $L$ for the partitions detected by LPAf and plotted them in Fig.~\ref{L_LPAf_GT}. Due to the global minimum of $L$ is not available, $L$-values of the planted partitions are adopted as references. As can be seen, when $\mu$ is small, i.e., the community structure is clear enough, the detected partitions and the true partitions almost have the same values of $L$, which indicates that LPAf correctly finds the real communities in the corresponding range of $\mu$. When $\mu$ increases to a specific value, $L$ decreases rapidly to a stable value on smaller networks, which corresponds to the trivial partition that the whole network is regarded as a single community. As the trivial partition has a lower value of $L$ than the planted one above a certain value of $\mu$, LPAf cannot detect any non-trivial communities within this range. However in larger networks, LPAf yields larger $L$ than that of the planted partition above a certain value of $\mu$, which implies that LPAf is trapped in a suboptimal valley in $L$ space.

\subsection{Tests on real-world networks}
\label{real_world_results}
We also applied the algorithms to several real-world networks that are commonly used for tests. The details of such networks are listed in Table \ref{networks}.
\begin{table}
	\centering
	\caption{\newline Real-world networks with community structure.}
	\label{networks}
	\resizebox{0.49\textwidth}{!}{
		\begin{tabular}{llll}
			\hline\noalign{\smallskip}
			Network & Reference & Vertices & Edges \\
			\noalign{\smallskip}\hline\noalign{\smallskip}
			karate & Zachary's karate club \cite{10.2307/3629752} & 34 & 78 \\
			dolphins & Dolphin social network \cite{Lusseau2003} & 62 & 159 \\
			books & Books about US politics \cite{Krebs2008} & 105 & 441 \\
			football & American College football \cite{doi:10.1137/S003614450342480} & 115 & 613 \\
			blogs & Political blogs \cite{Adamic:2005:PBU:1134271.1134277} & 1490 & 16715 \\
			netsci & Network scientists \cite{PhysRevE.74.036104} & 1589 & 2742 \\
			power & US power grid \cite{Watts1998} & 4941 & 6594 \\
			mat-cond & Condensed matter collaborations \cite{Newman2001} & 16726 & 47594 \\
			\hline
		\end{tabular}}
	\end{table}
	
We first compared directly the stability of different methods. All the methods are applied to each network 1000 times and the numbers of distinct detected partitions are reported. The pairwise VOI of the partitions are also computed to further evaluate the robustness of the methods. FineTune is not considered here since it is a deterministic algorithm. Due to the time complexity, two larger networks, \emph{power} and \emph{mat-cond}, are excluded from the analysis. Results are shown in Tables \ref{Distinct} and \ref{PairwiseVOI}. It is shown that LPAf is comparatively stable with less distinct partitions in most cases. LPA and Infomap are relatively unstable, even on smaller networks. Louvain method and nsdLPA have similar robustness, except on the \emph{netsci} network where Louvain method yields the most stable results. Moreover, as shown in Table \ref{PairwiseVOI}, the values of pairwise VOI between the partitions revealed by LPAf are lower than those for other methods in most cases. This concludes that LPAf is significantly more robust than LPA, and performs fairly stable.
\begin{table}
	\caption{Analysis of the stability of different methods. We report the number of distinct community structures obtained over 1000 runs.}
	\label{Distinct}
	\resizebox{0.49\textwidth}{!}{
		\begin{tabular}{llllll}
			\hline\noalign{\smallskip}
			Network & LPA & nsdLPA & LPAf & Louvain & Infomap \\
			\noalign{\smallskip}\hline\noalign{\smallskip}
			karate & 81 & 11 & \textbf{9} & 23 & 32  \\
			dolphins & 425 & 52 & 72 & \textbf{39} & 609 \\
			books & 191 & 75 & \textbf{10} & 73 & 725 \\
			football & 464 & 78 & \textbf{33} & 47 & 706 \\
			netsci & 1000 & 1000 & 496 & \textbf{181} & 1000 \\
			\hline
		\end{tabular}}
	\end{table}
	
	\begin{table}
		\caption{Analysis of the stability of different methods. We report the average pairwise VOI of the corresponding partitions obtained over 1000 runs.}
		\label{PairwiseVOI}
		\resizebox{0.49\textwidth}{!}{
			\begin{tabular}{llllll}
				\hline\noalign{\smallskip}
				Network & LPA & nsdLPA & LPAf & Louvain & Infomap \\
				\noalign{\smallskip}\hline\noalign{\smallskip}
				karate & 0.5189(4) & 0.2269(2) & \textbf{0.00482(2)} & 0.1967(2) & 0.2021(2)  \\
				dolphins & 0.4308(2) & \textbf{0.1387(1)} & 0.2130(1) & 0.2089(2) & 0.3177(1) \\
				books & 0.2989(1) & 0.1803(1) & \textbf{0.03818(9)} & 0.1677(1) & 0.3095(1) \\
				football & 0.14251(9) & 0.05192(4) & \textbf{0.02157(4)} & 0.05211(6) & 0.12204(8) \\                                                                                                                                                                                                                                                                                                                                                                                                                                                                                                                                      
				netsci & 0.037384(6) & 0.027995(5) & 0.008604(5) & \textbf{0.006858(5)} & 0.018963(6) \\
				\hline
			\end{tabular}}
		\end{table}

\begin{figure}
	\centering
	\resizebox{0.48\textwidth}{!}{%
		\includegraphics{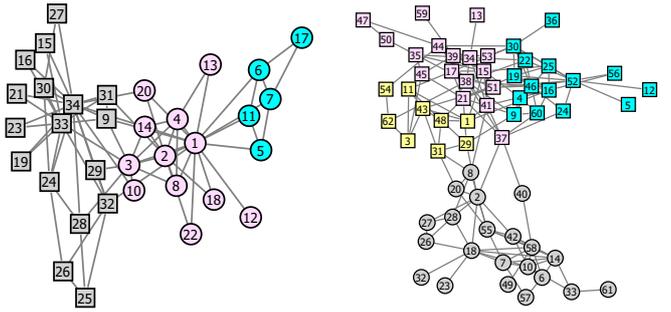}
	}
	\caption{(Color online) Communities detected by LPAf on \emph{karate} (left panel) and \emph{dolphin} (right panel) networks. The detected communities are distinguished by their colors, whereas the actual communities are represented by node shapes.}
	\label{karate_dolphins}
\end{figure}

Next, we detailedly analyzed the three networks (\emph{karate}, \emph{dolphins}, and \emph{football}) which have known community structures. Fig.~\ref{karate_dolphins} shows the communities detected by LPAf on \emph{karate} and \emph{dolphin} networks with the lowest $L$. Zachary's karate club is a social network of friendships between 34 members of a karate club at a US university in the 1970s. It splits into two smaller clubs after a dispute between club president John (node 34) and instructor Mr. Hi (node 1). As can be seen, three communities are discovered in this network by our algorithm. One of the two real communities is divided into two small ones (as shown in Fig.~\ref{karate_dolphins} (left panel)). The dolphin social network describes the frequent associations between 62 dolphins living off Doubtful Sound, New Zealand. The links represent that dolphins are observed to stay together more often than expected by chance during the years from 1994 to 2001. Four communities identified by our algorithm in this network are shown in Fig.~\ref{karate_dolphins} (right panel).

\begin{figure}
	\centering
	\resizebox{0.48\textwidth}{!}{%
		\includegraphics{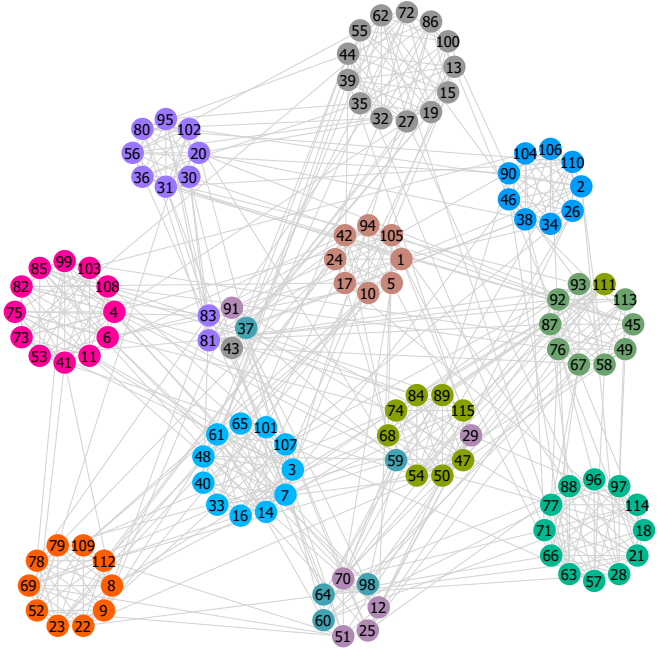}
	}
	\caption{(Color online) The \emph{football} network with each node representing a NCAA team, and each edge denoting a game played in 2000 between two teams they co-join. Colors represent the communities detected by LPAf. The 12 NCAA conferences are grouped into circles.}
	\label{football}       
\end{figure}

The \emph{football} network describes football games among Division IA colleges during regular season Fall 2000. As shown in Fig.~\ref{football}, the 115 nodes in the network represent teams, which are grouped into eleven different conferences, except for five independent teams. The regular season games between each pair of teams are shown as 613 edges of the network. Our algorithm identifies eleven communities within this network, as shown in Fig.~\ref{football}. Among them, eight conferences are correctly identified. The three remaining communities closely resemble the Conference USA, Sun Belt and Western Athletic conferences. Five independent teams that do not belong to any conference tend to be grouped with the conferences which they are most closely associated.

For comparison, we applied different methods to \emph{karate}, \emph{dolphins}, \emph{books} and \emph{football} networks, and measured the NMI between the real partitions and those detected by different methods. The average values of NMI over 1000 runs are shown in Table~\ref{averageNMI}. FineTune is deterministic and thus we only run it once. As one can see, LPAf performs fairly well on the \emph{karate} and the \emph{football} networks, although not the best. However it does not work well on the other two networks. The reason could be that the known partitions of these two networks do not have the lowest values of $L$, which prevents LPAf from detecting the real communities on these networks.
\begin{table}
	\caption{Average NMI between the real communities and those identified by the algorithms. Results are averages over 1000 different runs.}
	\label{averageNMI}
	\resizebox{0.49\textwidth}{!}{
		\begin{tabular}{lllllll}
			\hline\noalign{\smallskip}
			Network & LPA & nsdLPA & LPAf & Infomap & Louvain & FineTune \\
			\noalign{\smallskip}\hline\noalign{\smallskip}
			karate & 0.689(7) & \textbf{0.833(3)} & 0.821(1) & 0.751(2) & 0.651(1) & 0.5925 \\
			dolphins & \textbf{0.622(3)} & 0.606(1) & 0.520(1) & 0.506(1) & 0.493(1) & 0.4338 \\
			books & \textbf{0.5494(9)} & 0.5395(7) & 0.5391(3) & 0.5414(9) & 0.5421(7) & 0.4146 \\
			football & 0.8834(9) & 0.9039(3) & 0.9197(3) & 0.8994(6) & 0.8787(5) & \textbf{0.9242} \\
			\hline
	\end{tabular}}
\end{table}

In Table \ref{averagemodularity}, we also reported average modularity $Q$ of the detected partitions for all networks so as to enable a complete comparison. It is not surprising that Louvain method yields the highest values on almost all networks, because it is based on the optimization of modularity. Therefore, for clarity, we show the results of Louvain method in the rightmost column of the table. We also mark the best results of the rest methods in bold type. As one can see, LPAf achieves the best performance among those methods which do not directly optimize modularity in most cases.
\begin{table}
	\caption{Average modularity $Q$ of partitions identified by different algorithms. Results are the averages over 1000 different runs. Except the Louvain method, the best values are marked as boldface.} 
	\label{averagemodularity}
	\resizebox{0.49\textwidth}{!}{
		\begin{tabular}{llllllll}
			\hline\noalign{\smallskip}
			Network & True & LPA & nsdLPA & LPAf & Infomap & FineTune & Louvain \\
			\noalign{\smallskip}\hline\noalign{\smallskip}
			karate & 0.3718 & 0.344(3) & 0.3747(2) & 0.4008(1) & 0.3994(2) & \textbf{0.4174} & 0.4154(2) \\
			dolphins & 0.3787 & 0.482(1) & \textbf{0.5239(1)} & 0.5216(2) & 0.5067(5) & 0.4547 & 0.5206(1) \\
			books & 0.4149 & 0.4959(5) & 0.5183(2) & \textbf{0.52641(6)} & 0.5163(2) & 0.4855 & 0.52626(6) \\
			football & 0.554 & 0.5893(4) & 0.5673(5) & \textbf{0.60052(4)} & 0.5907(2) & 0.6005 & 0.60402(5) \\
			netsci &  & 0.9028(1) & 0.9093(1) & \textbf{0.9314(1)} & 0.9313(2) & 0.7641 & 0.95904(2) \\
			power &  & 0.7175(4) & 0.7204(3) & 0.8295(2) & \textbf{0.8297(2)} & 0.6036 & 0.93584(7) \\
			mat-cond &  & 0.7167(3) & 0.7270(2) & 0.7695(1) & \textbf{0.7758(2)} & 0 & 0.8479(1) \\
			\hline
	\end{tabular}}
\end{table}

In Table \ref{Qdsbar}, we presented average modularity density $Q_{ds}$ of partitions detected by different methods. FineTune is based on the optimization of modularity density. Therefore, we show the results of FineTune in the last column of the table and highlight the best results of the rest methods in bold type. As one can see, LPAf performs quite well in terms of $Q_{ds}$ in most cases.
\begin{table}
	\centering
	\caption{Average modularity density $Q_{ds}$ of partitions identified by different algorithms. Results are the averages over 1000 different runs. Except the FineTune method, the best values are marked as boldface.}
	\label{Qdsbar}
	\resizebox{0.49\textwidth}{!}{
		\begin{tabular}{llllllll}
			\hline\noalign{\smallskip}
			Network & True & LPA & nsdLPA & LPAf & Infomap & Louvain & FineTune \\
			\noalign{\smallskip}\hline\noalign{\smallskip}
			karate & 0.1823 & 0.197(2) & 0.1849(7) & 0.2168(1) & 0.2164(8) & \textbf{0.2284(3)} & 0.231 \\
			dolphins & 0.1362 & 0.184(2) & 0.1967(7) & \textbf{0.2060(5)} & 0.196(1) & 0.2009(9) & 0.264 \\
			books & 0.1267 & 0.174(1) & 0.1952(7) & \textbf{0.1986(3)} & 0.193(1) & 0.1972(4) & 0.2506 \\
			football & 0.4281 & 0.432(3) & 0.465(2) & \textbf{0.482(2)} & 0.457(2) & 0.437(2) & 0.4909 \\
			netsci &  & \textbf{0.6417(6)} & 0.6409(4) & 0.6136(4) & 0.6093(5) & 0.5029(3) & 0.4866 \\
			power &  & 0.2309(3) & \textbf{0.2339(3)} & 0.1527(2) & 0.1462(2) & 0.02067(7) & 0.3106 \\
			mat-cond &  & \textbf{0.3036(3)} & 0.2979(3) & 0.2526(1) & 0.2401(2) & 0.07047(9) & 0.0003 \\
			\hline
	\end{tabular}}
\end{table}

In Table \ref{averageL}, we compared different methods in terms of $L$. $L$-values of the true partitions are presented as references. As seen, LPAf achieves the best performance in most cases. It should be pointed out that the true partitions do not possess the global minimum of $L$. LPAf always obtains a lower $L$ than that of the true partition in some networks. This explains why LPAf cannot detect the real communities correctly on these networks.
\begin{table}
	\caption{Average description length $L$ of partitions identified by different algorithms. Results are the averages over 1000 different runs.} 
	\label{averageL}
	\resizebox{0.49\textwidth}{!}{
		\begin{tabular}{llllllll}
			\hline\noalign{\smallskip}
			Network & True & LPA & nsdLPA & LPAf & Infomap & Louvain & FineTune \\
			\noalign{\smallskip}\hline\noalign{\smallskip}
			karate & 4.3408 & 4.392 & 4.3483 & \textbf{4.2996(4)} & 4.319(1) & 4.418(1) & 4.4018 \\
			dolphins & 5.0786 & 5.068 & \textbf{4.9982(9)} & 5.099(1) & 5.159(2) & 5.095(1) & 5.7197 \\
			books & 6.0373 & 5.658(2) & 5.618(2) & \textbf{5.5875(7)} & 5.653(2) & 5.603(1) & 6.1893 \\
			football & 6.3784 & 6.091(3) & 6.311(4) & 6.0503(3) & 6.104(2) & \textbf{5.9811(4)} & 6.055 \\
			netsci &  & 4.135(1) & 4.061(1) & \textbf{3.7949(4)} & 3.8142(7) & 3.9716(5) & 6.7201 \\
			power &  & 8.467(1) & 8.417(1) & \textbf{6.8032(6)} & 6.8549(6) & 7.348(1) & 10.4736 \\
			mat-cond &  & 9.419(4) & 9.239(3) & \textbf{8.497(1)} & 8.608(2) & 9.125(2) & 13.4179 \\
			\hline
	\end{tabular}}
\end{table}

\begin{figure}
	\centering
	\resizebox{0.48\textwidth}{!}{%
		\includegraphics{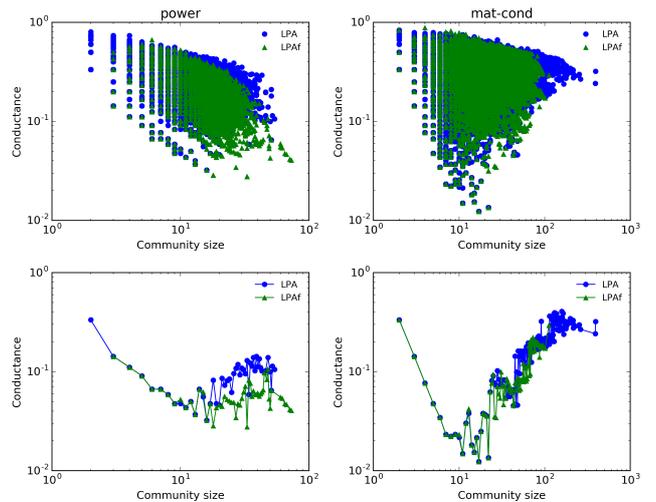}
	}
	\caption{(Color online) Comparison of LPAf and LPA on \emph{power} and \emph{mat-cond} networks. The conductances of individual communities (top panel), and the minimum conductances (bottom panel) at different scales are presented. Results were obtained over 100 runs.}
	\label{conductance_power_condmat}
\end{figure}

Lastly, we further analyzed the two larger networks, \emph{power} and \emph{mat-cond}. For simplicity, we only compared LPA and LPAf. We ran each method 100 times and analyze the conductances of the detected communities at various scales. The results are given in the form of NCP plots, as shown in Fig.~\ref{conductance_power_condmat}. NCP plots evaluate the quality of the best community (in terms of conductance) as a function of its size. Previous studies show that many kinds of real-world networks exhibit a common characteristic structure of NCP plots, i.e., initial decreasing and subsequent increasing trend \cite{leskovec2009community}.

In the case of \emph{power} network, LPAf detects communities on a much boarder scale with significant lower conductances, including also larger communities with around 80 nodes. On the \emph{mat-cond} network, both LPAf and LPA find the best communities at the same scale (i.e, at around 15 nodes), while the conductances of LPAf are sightly lower than that of LPA. Note that LPA reveals a number of larger communities with significant high conductances in both networks (i.e., blue circles in the top right part of the top two plots of Fig.~\ref{conductance_power_condmat}), which could be that many tie-breaks encountered in the label propagation process contributes to the formation of some large communities with high conductances.

\subsection{Time complexity}
Given a network with $n$ nodes and $m$ edges, let $k$ be the maximum degree of nodes in this network. The time complexity of each step of LPAf is roughly estimated as follows:
\begin{enumerate}
	\item \emph{Initialization} takes time of $O(n)$. Assigning a unique label to each node takes time of $O(n)$.
	\item \emph{Label propagation} takes time at most $O(nk)$. For each node, it iterates through at most $k$ neighbors, thus, the upper bound of cost time of this step is $O(nk)$.
	\item \emph{Merging communities} takes time at most $O(m\log n)$. Merging pairs of communities using MSG requires a time of $O(m\log n)$ in the worst case (see Ref.~\cite{Oades2008} for detailed analysis).
\end{enumerate}
Steps 2 and 3 are repeated, so the time per iteration is $O(kn+m\log n)$. Consequently, the time complexity of LPAf is roughly $O(kn+m\log n)$.

\begin{figure}
	\centering
	\resizebox{0.45\textwidth}{!}{%
		\includegraphics{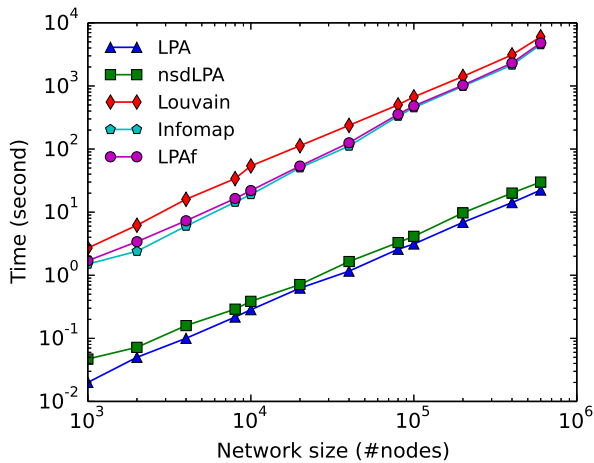}
	}
	\caption{(Color online) Comparison of running time for different algorithms on LFR networks with various sizes. Each data point is averaged over 30 different runs. The parameters of LFR networks are: the average degree is 20, the maximum degree is 50, the mixing parameter $\mu=0.4$ and the range of community sizes $C=[20,100]$.}
	\label{speedbench}       
\end{figure}

To evaluate the efficiency of LPAf, we have run LPA, nsdLPA, LPAf, Infomap and Louvain method on LFR networks with different sizes. Due to the high time complexity, FineTune is not considered in the benchmark situation. We repeated each experiment 30 times and reported the average running times. As shown in Fig.~\ref{speedbench}, the time complexity of LPA and nsdLPA is quite lower compared to the rest three methods. Still, all methods exhibit near linear time complexity and can be easily scaled to larger networks.

\section{Conclusion}
\label{conclu}
In this paper, we propose a modified label propagation algorithm (LPAf) to detect community structures in networks. In this algorithm, we introduce a new update rule which updates the label of a node by compressing a description of probability flow. Besides, by employing a multi-step greedy agglomerative algorithm, we merge pairs of communities so as to escape local minima in $L$-space. Furthermore, an incomplete update condition is adopted to accelerate the convergence.

We test the proposed algorithm on both synthetic and real-world networks, and compare its performance with that of the other five widely used methods in terms of \emph{modularity}, \emph{modularity density}, NMI, VOI and \emph{conductance}. Firstly, we find that LPAf performs very well on synthetic networks. In contrast to the Louvain method, LPAf is able to detect small communities in large networks. Secondly, we find that, LPAf detects communities which have lower conductances than that of LPA; by minimizing $L$, LPAf may fail to detect the real community structure which does not have the lowest $L$; LPAf is generally more stable than LPA. Finally, we analyze the time complexity of LPAf and find that it depends linearly on the network size in sparse networks.

In the future work, we intend to test our algorithm on weighted and directed networks. We also plan to extend our approach to overlapping community detection by allowing each node possess multi-labels.

\section*{Acknowledgements}
\label{ack}
This work was in part supported by the Program of Introducing Talents of Discipline to Universities under grant no. B08033, and National Natural Science Foundation of China (Grant No. 11505071).

\section*{Author contribution statement}
J.H. designed the algorithm, implemented the experiments, and prepared all the figures. J.H., L.Z. and Z.S. analyzed the results. All authors wrote, reviewed and approved the manuscript.

\numberwithin{equation}{section}
\section*{Appendix A: The change of average description length when a node moves from one community to another}
\label{appendixa}
From Eq.~(\ref{expandedmapequation}), for undirected and unweighted networks, the change of average description length when a node $\alpha$ updates its label from $i$ to $j$ is given by,
\begin{equation}
\small
\begin{split}
\Delta L(\alpha, i, j) =& \left(q_\curvearrowright+\delta q_\curvearrowright\right)\log\left(q_\curvearrowright+\delta q_\curvearrowright\right) - q_\curvearrowright\log\left(q_\curvearrowright\right) \nonumber \\
&-2\left[\left(q_{i\curvearrowright}+\delta q_{i\curvearrowright}\right)\log\left(q_{i\curvearrowright}+\delta q_{i\curvearrowright}\right) - q_{i\curvearrowright}\log\left(q_{i\curvearrowright}\right)\right] \nonumber \\
&-2\left[\left(q_{j\curvearrowright}+\delta q_{j\curvearrowright}\right)\log\left(q_{j\curvearrowright}+\delta q_{j\curvearrowright}\right) - q_{j\curvearrowright}\log\left(q_{j\curvearrowright}\right)\right] \nonumber \\
&+ \left(p_{i\circlearrowright}+\delta p_{i\circlearrowright}\right)\log\left(p_{i\circlearrowright}+\delta p_{i\circlearrowright}\right) - p_{i\circlearrowright}\log\left(p_{i\circlearrowright}\right) \nonumber \\
&+ \left(p_{j\circlearrowright}+\delta p_{j\circlearrowright}\right)\log\left(p_{j\circlearrowright}+\delta p_{j\circlearrowright}\right) - p_{j\circlearrowright}\log\left(p_{j\circlearrowright}\right)
\end{split}
\end{equation}
with
$$\delta q_\curvearrowright = \delta q_{i\curvearrowright} + \delta q_{j\curvearrowright},$$
$$\delta p_{i\circlearrowright} = \delta q_{i\curvearrowright} - p_\alpha,$$
$$\delta p_{j\circlearrowright} = \delta q_{j\curvearrowright} + p_\alpha,$$
$$\delta q_{i\curvearrowright} = \sum_{\beta\in\partial \alpha \cap V_i}\frac{1}{2m} - \sum_{\beta\in\partial \alpha \setminus V_i}\frac{1}{2m},$$
$$\delta q_{j\curvearrowright} = \sum_{\beta\in\partial \alpha \setminus V_j}\frac{1}{2m} - \sum_{\beta\in\partial \alpha \cap V_j}\frac{1}{2m},$$
where $m$ is the total number of edges of the network, $V_i$ and $V_j$ are the nodes in community $i$ and $j$ respectively, and $\partial\alpha$ is the neighbors of $\alpha$. Extension to directed and weighted networks is straightforward.

\bibliographystyle{epj}
\bibliography{LPAf}

\end{document}